\documentclass[]{raa}            

\usepackage{amsmath,amssymb,amsxtra,amsfonts}
\usepackage{graphicx,times}
\usepackage{epsfig}
\usepackage{natbib}
\usepackage{txfonts}

\def\etal{{\frenchspacing\it et al.~}}
\def\ie{{\frenchspacing\it i.e.~}}
\def\eg{{\frenchspacing\it e.g.~}}
\def\etc{{\frenchspacing\it etc.~}}
\newcommand{\ud}{\mathrm{d}}

\def\Om{\Omega_{\rm{m}}}
\def\Ol{\Omega_{\Lambda}}
\def\Ok{\Omega_{\rm{K}}}
\def\Obh{\Omega_{\rm{b}}h^2}
\def\Ob{\Omega_{\rm{b}}}
\def\fgas{\hbox{$f_{\rm{gas}}~$}}
\def\fgasz{\hbox{$f_{\rm{gas}}(z)$} }
\def\r2500{\hbox{$R_{\Delta=2500}~$}}
\def\rr500{\hbox{$R_{\Delta=500}~$}}
\newcommand{\dl}{D_{\rm{L}}}
\newcommand{\da}{D_{\rm{A}}}
\newcommand{\dlz}{D_{\rm{L}}(z)}
\newcommand{\daz}{D_{\rm{A}}(z)}
\newcommand{\dls}{D^{\ast}_{\rm{L}}(z)}
\newcommand{\das}{D^{\ast}_{\rm{A}}(z)}
\newcommand{\as}{\alpha_{\rm{s}}}
\newcommand{\ab}{\alpha_{\rm{b}}}
\newcommand{\asz}{\alpha_{\rm{s}}z}
\newcommand{\abz}{\alpha_{\rm{b}}z}

\newcommand{\g}{\gamma}

\def\chandra{\it Chandra\rm~}
\def\ovro{\it OVRO\rm~}
\def\bima{\it BIMA\rm~}
\def\hst{\it Hubble Space Telescope\rm~}

\begin{document}

   \title{Testing X-ray Measurements of Galaxy Cluster Gas Mass Fraction Using the Cosmic Distance-Duality Relation}

   \volnopage{Vol.0 (200x) No.0, 000--000}      
   \setcounter{page}{1}          

   \author{Xin Wang
      \inst{1,2,3}
   \and Xiao-Lei Meng
      \inst{2,3}
   \and Y. F. Huang
      \inst{1}
   \and Tong-Jie Zhang
      \inst{2,4}
   }

   \institute{School of Astronomy and Space Science, Nanjing University, Nanjing 210093, China; {\it hyf@nju.edu.cn}\\
        \and
             Department of Astronomy, Beijing Normal University, Beijing 100875, China\\
        \and
             National Astronomical Observatories, Chinese Academy of Sciences, Beijing
100012, China\\
        \and
             Kavli Institute for Theoretical Physics China, CAS, Beijing 100190, China\\
   }

   \date{Received~~year month day; accepted~~year~~month day}

\abstract{
We propose a consistency test of some recent X-ray gas mass fraction ($f_{\rm{gas}}$) measurements in galaxy clusters, using the cosmic distance-duality relation,
$\eta_{\rm{theory}}=\dl(1+z)^{-2}/\da$, with luminosity distance ($\dl$) data from the
Union2 compilation of type Ia supernovae. We set $\eta_{\rm{theory}}\equiv1$, instead of
assigning any redshift parameterizations to it, and constrain the cosmological
information preferred by $f_{\rm{gas}}$ data along with supernova observations.
We adopt a new binning method in the reduction of the Union2 data, in order to minimize the
statistical errors. Four data sets of X-ray gas mass fraction, which are reported by Allen et al. (2 samples), LaRoque et al. and Ettori et al., are detailedly analyzed against two theoretical modelings of $f_{\rm{gas}}$. The results from the analysis of Allen et al.'s samples prove the feasibility of our method. It is found that the preferred cosmology by LaRoque et al.'s sample is consistent with its reference cosmology within 1-$\sigma$ confidence level. However, for Ettori et al.'s $f_{\rm{gas}}$ sample, the inconsistency can reach more than 3-$\sigma$ confidence level and this dataset shows special preference to an $\Ol=0$ cosmology.
}

   \authorrunning{Wang et al. }            
   \titlerunning{Using the DD Relation to constrain \fgas measurements of galaxy clusters}  

\maketitle

\section{Introduction}\label{sect:intro}

As the largest virialized objects, clusters of galaxies play a critical role in promoting
our knowledge of matter distributions in distant universe as well as the formation and
evolution of large-scale structures \citep{2005RvMP...77..207V,2011ARA&A..49..409A}.
Using galaxy clusters, there have been accumulated work to obtain the Hubble constant
\citep{mason01,cunha07}, to put constraints on the matter/energy content of the universe
\citep{2003PhRvD..68b3510L,2009ApJ...692.1060V}, to study the evolution of underlying
massive halos via N-body and hydrodynamical simulations
\citep{1998ApJ...503..569E,2005ApJ...625..588K}, and to measure distance scales independent
of cosmological models using clusters as standard rulers
\citep{2005ApJ...625..108D,2006ApJ...647...25B}. In practice, the observed abundance of
high redshift ($z\sim1$) massive clusters provides strong indirect evidence for the
existence of dark energy \citep{1998ApJ...504....1B}, which is firstly introduced to
explain the cosmic acceleration based on the observations of supernovae type Ia (SNe
Ia)\citep{1998AJ....116.1009R,1999ApJ...517..565P}.

The cluster gas mass fraction measured from X-ray observations, $\fgas=M_{\rm
gas}/M_{\rm tot}$, \ie the ratio between the mass of intracluster medium (ICM) gas and
the total mass of the cluster, serves as a powerful cosmological probe. According to
\citet{1993Natur.366..429W}, the baryon budget of rich clusters should reflect the cosmic
value of $\Ob/\Om$, where $\Ob$ and $\Om$ are the mean baryonic and total matter
densities of the universe, in units of the critical density, $\rho_{\rm
c}(z)=3H(z)^2/(8\pi G)$. Moreover, the estimates from \citet{1998ApJ...503..518F}
indicate that the baryon mass constituent of clusters is dominated by hot intracluster
gas, with the contribution from optically luminous stellar component less than twenty
percent, and other sources negligible. In a series of work, using \fgas as a
proxy of cosmic baryon budget, Allen \etal (2002, 2004, 2008) improved the analysis
method, enlarged the cluster sample (from 7 to 26, then to 42 data points),
and tightened the constraints on cosmological parameters. The idea of determining dark
energy equation of state is also explored in \citet{2010MNRAS.406.1759M}, via the combination
of \fgas measurements and other observations. \citet{2003MNRAS.342..287A} make use of \fgas to
constrain the relation between the normalization of power spectrum of mass fluctuations, \ie $\sigma_8$, and $\Om$.
\citet{2006MNRAS.365.1021E} investigated how miscellaneous physical
processes in clusters, \eg radiative cooling, star formation activities and galactic wind
feedback, affect the baryon fraction measurements, through hydrodynamical simulations.

In calculating \fgas , a general duality between two distance scales,
\begin{equation}\label{eq:da-dl}
\eta_{\rm{theory}}=\frac{\dl}{\da}(1+z)^{-2}=1,
\end{equation}
is assumed in almost all previous studies \citep[\eg see][footnote
1]{2008MNRAS.383..879A}, where $\dl$ and $\da$ stand for luminosity and angular diameter
distances. This distance duality was firstly proposed by \citet{1933PMag...15..761E}, and
usually termed as the Etherington's reciprocity relation or the cosmic distance-duality relation (CDDR). The CDDR is vital for
observational cosmology, since any marked intrinsic violation of the CDDR may give rise
to exotic physics \citep{2004PhRvD..69j1305B}. The validity of the CDDR only depends on photon conservation on cosmic scales and the condition that the effect of gravitational lensing should be negligible, regardless of any metric theory of gravities. Several research groups have used various observational data to test the validity of the CDDR \citep{uzan04,deb06,av10,2011arXiv1104.2497L}. Especially, using galaxy clusters' $\da$ from the joint analysis of X-ray surface brightness and Sunyaev-Zel'dovich technique and SNe Ia's $\dl$ from Union
compilation, \citet{2010ApJ...722L.233H} performed a cosmologically independent test on the
CDDR. Following this route, \citet{2011ApJ...729L..14L} tested the CDDR using the latest
compilation comprised of 557 SNe Ia \citep[Union2 compilation,][]{2010ApJ...716..712A}.
Both \citet{2010ApJ...722L.233H} and \citet{2011ApJ...729L..14L} employed a moderate redshift
criterion, $\Delta z=| z_{\rm cluster} - z_{\rm SN}| <0.005$, to select the nearest SN Ia
for every galaxy cluster. \citet{lp} improved this analysis by developing two
sophisticated methods to guarantee all appropriate SNe Ia data selected, so as to reduce
statistical errors. They found that the CDDR is compatible with the elliptically modeled
galaxy cluster sample \citep{2005ApJ...625..108D} at $1\sigma$ confidence level (CL).
However for some parameterizations, the CDDR can not be accommodated even at $3\sigma$
CL for the spherical $\beta$-model cluster sample \citep{2006ApJ...647...25B}. Therefore
their results support that the marked triaxial ellipsoidal model is a better hypothesis
describing the structure of the galaxy cluster compared with the spherical $\beta$ model,
if the CDDR holds valid in cosmological observations. \citet{2012A&A...538A.131H} has
arrived at similar conclusions.

More recently, \citet{2012MNRAS.420L..43G} proposed the idea of testing the validity of the CDDR using X-ray \fgas data. In obtaining \fgas sample, one has to assume some reference cosmology to solve \fgas dependence upon metric distances. In consequence, this test for the CDDR is not cosmology independent. Moreover, because the CDDR is already assumed valid in the measurements of \fgas, this test is not observationally robust. In this paper, we reverse the procedure of \citet{2012MNRAS.420L..43G}, via fixing $\eta_{\rm{theory}}\equiv1$ instead of assigning any redshift parameterizations to $\eta_{\rm{theory}}$ \citep[see][Eq.(15)]{2012MNRAS.420L..43G}, and then constrain the preferred cosmological
information by a given set of \fgas data. Thus a straightforward comparison between the \fgas sample preferred cosmology and its reported reference model is allowed. This may be a viable approach to present a consistency test of current \fgas measurements.

This paper is organized as follows. In Section~2, we briefly review the theoretical basis of formulating \fgas as a function of redshift and metric distances. The data samples and analysis method are then described in Section~3. Section~4 presents the main results, and Section~5 gives the conclusions and discussion.

\section{Theory: incorporating the CDDR into gas fraction}\label{sect:theo}

The possibility of deriving cosmological constraints through the apparent redshift
dependence of cluster baryon mass fraction was firstly discussed by
\citet{1996PASJ...48L.119S} and \citet{1997NewA....2..309P}. Supposing X-ray emission
from ICM gas is mainly due to thermal bremsstrahlung \citep{1988xrec.book.....S}, the gas
mass enclosed within a measurement radius $R$ can be derived as,
\begin{eqnarray}\label{eq:mgas}
M_{\rm{gas}}(<R)&=&\left[\frac{3\pi\hbar{m_{\rm{e}}}c^2}{2(1+X)e^6}\right]^{1/2}\left(\frac{3m_{\rm{e}}c^2}{2\pi{k_{\rm{B}}}T_{\rm{e}}}\right)^{1/4}m_{\rm{H}}\nonumber\\
&&\mbox{\hspace{-1.5cm}}\times\frac{1}{[\overline{g_{\rm{B}}}(T_{\rm{e}})]^{1/2}}r_{\rm{c}}^{3/2}\left[\frac{I_{\rm{M}}(R/r_{\rm{c}},\beta)}{I_{\rm{L}}^{1/2}(R/r_{\rm{c}},\beta)}\right][L_{\rm{X}}(<R)]^{1/2}\;,
\end{eqnarray}
where $L_{\rm{X}}(<R)$ is the X-ray bolometric luminosity, $r_{\rm{c}}$ denotes the core
radius, and the other symbols have their usual meanings. Furthermore, under the
assumption of hydrostatic equilibrium and isothermality ($T_{\rm{e}}=\rm{const.}$) for
ICM, the total mass in a cluster of galaxies within $R$ is given by
\begin{equation}\label{eq:mtot}
M_{\rm{tot}}(<R)=-\left.\frac{k_{\rm{B}}T_{\rm{e}}R}{G\mu{m_{\rm{H}}}}\frac{\ud\ln{n_{\rm{e}}(r)}}{\ud\ln{r}}\right|_{r=R}.
\end{equation}

In the above estimations, the measurement radius is determined by fixing a certain value
for the cluster overdensity
($\Delta=3M_{\rm{tot}}(<R_{\Delta})/(4\pi\rho_{\rm{c}}(z_{\rm{cluster}})R_{\Delta}^3)$),
where $z_{\rm{cluster}}$ represents the cluster's redshift. Usually $\Delta$ is adopted
as 2500 \citep{2004MNRAS.353..457A,2006ApJ...652..917L} or 500
\citep{2009A&A...501...61E}. Discussion has been raised regarding which value is more
trustworthy in measuring \fgas\citep{2006ApJ...640..691V,2011ARA&A..49..409A}. We also
study this problem by analyzing two groups of \fgas datasets, assuming different values for
$\Delta$.

The reference cosmology enters these relations via
\begin{eqnarray}\label{eq:lxrc}
L_{\rm{X}}(<R)&=&4\pi\dl^2f_{\rm{X}}(<\theta),\\
r_{\rm{c}}&=&\theta_{\rm{c}}\da,\\
R&=&\theta\da.
\end{eqnarray}
Eqs. (\ref{eq:mgas}) and (\ref{eq:mtot}) then indicate
\begin{eqnarray}
M_{\rm{gas}}&\propto&\dl\da^{3/2},\\
M_{\rm{tot}}&\propto&\da.
\end{eqnarray}
Thus it is straightforward to derive
\begin{equation}\label{eq:fgasdlda}
\fgas=M_{\rm{gas}}/M_{\rm{tot}}\propto\dl\da^{1/2}.
\end{equation}
Note that in all previous measurements of \fgas, Eq. (\ref{eq:fgasdlda}) is
readily reduced to $\fgas\propto\da^{3/2}$, which already assumes the CDDR in the first
place, and therefore makes the test for the validity of the CDDR with \fgas data strongly biased. Aiming at using the CDDR to constrain \fgas samples, we employ more original forms for \fgas in subsequent analyses.

We model \fgas using the popular expression proposed by \citet{2004MNRAS.353..457A},
\begin{equation}\label{eq:fgas04}
\fgasz=\frac{b}{\left(1+0.19\sqrt{h}\right)}\cdot\frac{\Ob}{\Om}\cdot\left(\frac{\dls\das^{1/2}}{\dlz\daz^{1/2}}\right),
\end{equation}
with the dependence on metric distances modified according to Eq. (\ref{eq:fgasdlda}). A
more generalized form recently proposed by \citet{2008MNRAS.383..879A} is also
considered,
\begin{equation}\label{eq:fgas08}
\fgasz=\frac{K\g(b_0+b_1z)}{1+s_0(1+\asz)}\cdot\frac{\Ob}{\Om}\cdot\left(\frac{H(z)\daz}{H^{\ast}(z)\das}\right)^{\xi}\cdot\left(\frac{\dls\das^{1/2}}{\dlz\daz^{1/2}}\right),
\end{equation}
which has also been revised due to aforementioned intrinsic distance dependence.
In Eqs. (\ref{eq:fgas04}) and (\ref{eq:fgas08}), $\Ob$ stands for the baryonic
matter density, which can be inferred from the big bang nucleosynthesis. $b$ (or
$b(z)=b_0+b_1z$) represents the baryonic depletion independent (or dependent) of
redshift, as a consequence of the thermodynamical evolution of clusters. $h$ depicts the
Hubble constant via $H_0=100h$ km s$^{-1}$Mpc$^{-1}$, and is adopted from the final
results of the \hst Key Project \citep{2001ApJ...553...47F}. $s(z)=s_0\left(1+\asz\right)$
models the baryonic matter fraction contributed from stellar component. $\gamma$
considers the non-thermal pressure contributing to the hydrostatic equilibrium and
lowering $M_{\rm{tot}}$. $K$ stands for instrument calibration and $\xi$ corresponds to
the relationship between the characteristic radius and the angular aperture of
measurement. Table~1 summarizes two sets of a priori knowledge about these nuisance
parameters, for different \fgas samples measured under different $\Delta$.

Two sets of metric distances appear in Eqs. (\ref{eq:fgas04}) and (\ref{eq:fgas08}). The
distances with a mark of star correspond to the distances calculated from a certain
reference cosmological model, which in context of the $\Lambda$CDM cosmology are given by
\begin{eqnarray}\label{eq:dlsdas}
\dls=\das(1+z)^2&=&\frac{c(1+z)}{H_0}\frac{S(\omega)}{\sqrt{|\Ok|}},\\
\omega&=&\sqrt{|\Ok|}\int_0^z\frac{\ud\zeta}{E(\zeta)},\nonumber
\end{eqnarray}
where $S(\omega)$ is $\sinh(\omega),\omega,\rm{or}\sin(\omega)$ for $\Ok$ larger than,
equal to or smaller than zero, respectively.
$E(z)=\frac{H(z)}{H_0}=\left[\Om(1+z)^3+\Ok(1+z)^2+\Ol\right]^{1/2}$ represents the
$\Lambda$CDM expansion history. Usually, it is safe to write $\Om+\Ok+\Ol=1$, with $\Ol$
and $\Ok$ accounting for the constant dark energy density and the curvature of space. The
distances without the mark of star can be connected through the CDDR,
$\eta_{\rm{obs}}(z)=\frac{\dlz}{\daz (1+z)^2}$. Then we obtain
\begin{equation}\label{eta_obs04}
\eta_{\rm{obs}}(z)=\left(\frac{1+0.19\sqrt{h}}{b\Ob}\right)^2\cdot\frac{\Om^2}{\left(1+z\right)^6}\cdot\left(\frac{\dlz}{\das}\right)^3(\fgasz)^2
\end{equation}
for the \fgasz expression given by Eq. (\ref{eq:fgas04}), and
\begin{eqnarray}\label{eta_obs08}
\eta_{\rm{obs}}(z)&=&\left(\frac{K\g(b_0+b_1z)\Ob}{\left[1+s_0(1+\asz)\right]\Om}\right)^{\frac{2}{2\xi-1}}\cdot(1+z)^{\frac{-4\xi+6}{2\xi-1}}\nonumber\\
&&\cdot\left(\frac{H(z)}{H^{\ast}(z)}\right)^{\frac{2\xi}{2\xi-1}}\cdot\left(\frac{\das}{\dlz}\right)^{\frac{-2\xi+3}{2\xi-1}}(\fgasz)^{\frac{2}{-2\xi+1}}\;
\end{eqnarray}
for \fgasz given by Eq. (\ref{eq:fgas08}). In Eqs. (\ref{eta_obs04}) and (\ref{eta_obs08}),
$\das$ can be calculated according to Eq. (\ref{eq:dlsdas}). In order to obtain
$\eta_{\rm{obs}}(z)$, we still need the observational results of \fgasz and $\dlz$, which
are introduced in the next section.

\begin{table}
\begin{center}
\caption[]{Summary of the Priors and Systematic Allowances for Nuisance Parameters
Present in the Two $\eta_{\rm{obs}}(z)$ Expressions.\label{table:np}}
\begin{tabular}{lcc}
\hline\hline \noalign{\smallskip}
\multicolumn{3}{c}{$\eta_{\rm{obs}}(z)$ Expression of Eq. (\ref{eta_obs04})}
\tabularnewline \hline\noalign{\smallskip}
       &      \multicolumn{2}{c}{Allowance}             \\
Nuisance Parameter & A04 $\Lambda$CDM, A04 SCDM, L06 & E09, L06 ($\Delta=500$)      \\
\hline\noalign{\smallskip}
$\Obh$   &  $0.0214\pm0.0020$ &  --                     \\
$\Ob$    &  --                &  $0.0462\pm0.0012$      \\
$h$      &  $0.72\pm0.08$     &  $0.72\pm0.08$          \\
$b$      &  $0.824\pm0.089$   &  $0.874\pm0.023$        \\
\hline\noalign{\bigskip}

\multicolumn{3}{c}{$\eta_{\rm{obs}}(z)$ Expression of Eq. (\ref{eta_obs08})}
\tabularnewline \hline\noalign{\smallskip}
       &      \multicolumn{2}{c}{Allowance}             \\
Nuisance Parameter & L06  & E09, L06 ($\Delta=500$)            \\
\hline\noalign{\smallskip}
$K$    &  $1.0\pm0.1$       &  $1.0\pm0.1$              \\
$\g$   &  $(1.0,1.1)$       &  $(1.0,1.1)$              \\
$\Obh$ &  $0.0214\pm0.0020$ &  --                       \\
$\Ob$  &  --                &  $0.0462\pm0.0012$        \\
$h$    &  $0.72\pm0.08$     &  $0.72\pm0.08$            \\
$b_0$  &  $(0.65,1.0)$      &  $0.923\pm0.006$          \\
$b_1$  &  --                &  $0.032\pm0.010$          \\
$\ab$$~^a$  &  $(-0.1,0.1)$      &  --                  \\
$s_0$  &  $0.16\pm0.048$    &  $0.18\pm0.05$            \\
$\as$  &  $(-0.2,0.2)$      &  $(-0.2,0.2)$             \\
$\xi$  &  $0.214\pm0.022$   &  0.2$~^b$                 \\
\hline\hline \noalign{\smallskip}
\end{tabular}
\end{center}
\tablecomments{0.9\textwidth}{($a$) \citet{2008MNRAS.383..879A} used $b(z)=b_0(1+\abz)$ to stand for the depletion factor, where $b_0\times\ab$ is equivalent to $b_1$ in our definition (Eq. (\ref{eq:fgas08})). ($b$) This value is determined from Eq. (4) of Ettori et al. (2003).}
\end{table}

\section{Data Sets and Analysis Method}

Here we first describe the \fgas samples analyzed following the aforementioned idea and the
SNe Ia data that furnish $\dlz$. Then we describe as a whole the key procedures of our
method.

\subsection{Galaxy cluster Samples and SNe Ia Union2 Data}

\citet{2004MNRAS.353..457A} analyzed a sample of 26 luminous, dynamically relaxed galaxy
clusters observed with \chandra at redshift $0.07<z<0.9$. They used the NFW model
\citep{1997ApJ...490..493N} to parameterize cluster total mass profiles. Assuming
different reference cosmological models, \ie $\Lambda$CDM ($h=0.7,\Om=0.3,\Ol=0.7$) and
SCDM ($h=0.5,\Om=1,\Ol=0$) \citep[see][Table~2]{2004MNRAS.353..457A}, they provided with
two samples of \fgas, which are referred to as A04 $\Lambda$CDM and A04 SCDM,
respectively. For consistency, we only use Eq. (\ref{eta_obs04}) as $\eta_{\rm{obs}}(z)$
for these two samples, since they come from the same paper. The priors and systematic
allowances on nuisance parameters for these two samples can be found in
Table~\ref{table:np}. $\Delta=2500$ is chosen in measuring \fgas.

As a follow-up study of \citet{2003A&A...398..879E}, the paper by
\citet{2009A&A...501...61E} focused on 52 clusters of \chandra measurements, spanning
in the range of $0.3<z<1.3$. The electron density profiles are fit with a functional form
adapted from \citet{2006ApJ...640..691V}. We choose the dataset assuming a constant
temperature given by spectral analysis for each individual cluster
\citep[see][Table~1]{2009A&A...501...61E}, and quote this sample as E09 hereafter. Three
clusters with spectral temperatures below 4 keV are excluded. The reported reference
cosmology is $\Lambda$CDM ($h=0.7,\Om=0.3,\Ol=0.7$). The priors/allowances on nuisance
parameters are obtained mainly from the original paper by \citet{2009A&A...501...61E},
which fixes $\Delta=500$.

Combining \chandra X-ray observations and Sunyaev-Zel'dovich effect measurements from
\ovro/\bima interferometric arrays, \citet{2006ApJ...652..917L} obtained \fgas results of
38 massive galaxy clusters, in the redshift range $0.142-0.89$.  We use their X-ray \fgas
dataset assuming the gas distribution is described by the isothermal $\beta$-model
\citep{1976A&A....49..137C} with the central 100 kpc excised
\citep[see][Table~4]{2006ApJ...652..917L}. This sample also employs a reference cosmology
of $\Lambda$CDM ($h=0.7,\Om=0.3,\Ol=0.7$). The original sample (referred to as L06)
assumes $\Delta=2500$, and thus can be analyzed using the priors/allowances proposed by
Allen \etal (2004, 2008), since they adopt the same $\Delta$. Furthermore, we use the
correlation obtained by \citet{2006ApJ...640..691V},
$f_{\rm{gas},\Delta=2500}/f_{\rm{gas},\Delta=500}=0.84$, to derive a new \fgas sample at
$R_{\Delta=500}$, which is quoted as L06($\Delta=500$). Besides the errors contributed
from the original L06 data, a 10\% uncertainty is also added to the errors of
L06($\Delta=500$) data. For this sample, the priors/allowances are chosen to be exactly
the same as those for E09. Note that L06($\Delta=500$) is not directly measured at
$R_{\Delta=500}$. Its analysis result reflects the accuracy of \fgas measurement at
$R_{\Delta=2500}$.

To get the luminosity distances, we choose the Union2 compilation comprised of 557 SNe Ia
\citep{2010ApJ...716..712A}\footnote{Using the more updated Union2.1 compilation
\citep{2012ApJ...746...85S} does not influence our results. Comparing with the
Union2 sample, this dataset includes twenty-three new events over the high redshift
range ($0.6<z<1.4$), which has little overlap with the redshift ranges of the cluster samples.}.
The uncertainty of SNe Ia's absolute magnitude
\citep{2011ApJ...730..119R}, \ie a systematic error of 0.05 magnitudes, is also
considered as an additive covariance, and combined in quadrature among all distance
moduli, provided by the \textit{Supernova Cosmology
Project}\footnote{http://supernova.lbl.gov/Union/}. \citet{2012MNRAS.420L..43G}
used the criterion, $\Delta{z}=|z_{\rm{cluster}}-z_{\rm{SN}}|<0.006$, to select the
nearest SN Ia for each cluster for the sake of a direct test. However selecting merely
one SN Ia within a certain redshift range will definitely lead to larger statistical
errors \citep[see][footnote 7]{lp}. Instead, we take an inverse variance weighted
average of all the selected data,
\begin{equation}
\begin{array}{l}
\bar{\dl}=\frac{\sum\left(D_{\textrm{L}i}/\sigma^2_{D_{\textrm{L}i}}\right)}{\sum1/\sigma^2_{D_{\textrm{L}i}}},\\
\sigma^2_{\bar{\dl}}=\frac{1}{\sum1/\sigma^2_{D_{\textrm{L}i}}},
\end{array}\label{eq:dlsigdl}
\end{equation}
where $D_{\textrm{L}i}$ represents the $i$th selected luminosity distance within
$\Delta{z}<0.005$ and $\sigma_{D_{\textrm{L}i}}$ denotes its observational uncertainty.
What we ultimately utilize is $\bar{\dl}$, the weighted mean luminosity distance at the
corresponding $z_{\rm{cluster}}$, with $\sigma_{\bar{\dl}}$ being its uncertainty. This
binning method can significantly decrease statistical errors. Additionally, in all the
five \fgas samples, if a cluster is not associated with any SNe Ia within
$\Delta{z}<0.005$, then it is excluded to avoid large systematic uncertainties.

\subsection{Statistics}

Since it is assumed that $\eta_{\rm{theory}}\equiv1$, we can calculate $\chi^2$ as,
\begin{equation}\label{eq:chi2}
\chi^2=\sum\limits_z{\frac{\left(\eta_{\rm{obs}}(z)-1\right)^2}{\sigma_{\eta_{\rm{obs}}(z)}^2}},
\end{equation}
where $\eta_{\rm{obs}}(z)$ is given by Eq. (\ref{eta_obs04}) or (\ref{eta_obs08}), while
$\sigma_{\eta_{\rm{obs}}(z)}$ is obtained through error propagations from $\sigma_{\dlz}$
and $\sigma_{f_{\rm{gas}}(z)}$. The asymmetric uncertainties of L06 data are handled
using the technique proposed by \citet{2004physics...3086D}. The likelihood function,
$\mathbb{L}\propto\mathrm{e}^{-\chi^2/2}$, is calculated over a certain range of grids of
values for cosmological parameters, $\Om$ and $\Ol$. Then, after the marginalization over
nuisance parameters in Eq. (\ref{eta_obs04}) or (\ref{eta_obs08}), we can obtain the
posterior probability of each reference cosmological model.

For each \fgas sample, the marginalization process requires specific a priori knowledge
of all nuisance parameters. In our analysis, all the systematic allowances and priors,
listed in Table~1, are carefully chosen according to previous studies (Allen et al. 2004,
2008; Ettori et al. 2003, 2009). The best-fit values are defined as the marginalized
probability reaching its maximum. For 1-dimensional analysis giving constraint on the flat
$\Lambda$CDM reference cosmology (with only one parameter, $\Om$), the 1, 2 and
3$\sigma$ confidence levels (CLs) are defined with the marginalized probability
equivalent to $\rm{e}^{-1.0/2}$, $\rm{e}^{-4.0/2}$ and $\rm{e}^{-9.0/2}$ of the maximum,
whereas for the 2-dimensional constraint on $(\Om,\Ol)$, \ie on usual $\Lambda$CDM
cosmology, the ratios are taken to be $\rm{e}^{-2.30/2}$, $\rm{e}^{-6.17/2}$ and
$\rm{e}^{-11.8/2}$.

\begin{table}
\begin{center}
\caption[]{Summary of the Best-fit Values and 1$\sigma$ Uncertainties of the
Preferred Cosmological Parameters by Each \fgas Sample, given the CDDR and
Union2 SNe Ia Data.}
\begin{tabular}{lccc}
\hline\hline\noalign{\smallskip}
 & Flat $\Lambda$CDM & \multicolumn{2}{c}{General $\Lambda$CDM} \\
 & \multicolumn{1}{c}{\hrulefill} & \multicolumn{2}{c}{\hrulefill}     \\
Sample & $\Om$ & $\Om$ & $\Ol$ \\
\hline\noalign{\smallskip}

\multicolumn{4}{c}{$\eta_{\rm{obs}}(z)$ Expression of Eq. (\ref{eta_obs04})}

\tabularnewline \hline\noalign{\smallskip}
A04 $\Lambda$CDM & $0.282_{-0.060}^{+0.072}$ & $0.25_{-0.08}^{+0.14}$ & $0.52_{-0.52}^{+0.42}$   \\
A04 SCDM         & $0.545_{-0.112}^{+0.154}$ & $0.30_{-0.10}^{+0.17}$ & ${0.00_{-0.00}^{+0.20}}$ \\
E09              & $0.439_{-0.023}^{+0.026}$ & $0.38_{-0.03}^{+0.09}$ & ${0.00_{-0.00}^{+0.78}}$ \\
L06              & $0.284_{-0.052}^{+0.069}$ & $0.28_{-0.09}^{+0.15}$ & $0.69_{-0.44}^{+0.32}$      \\
L06($\Delta=500$)& $0.295_{-0.012}^{+0.013}$ & $0.31_{-0.04}^{+0.03}$ & $0.92_{-0.57}^{+0.32}$      \\

\hline\noalign{\smallskip}

\multicolumn{4}{c}{$\eta_{\rm{obs}}(z)$ Expression of Eq. (\ref{eta_obs08})}

\tabularnewline \hline\noalign{\smallskip}
E09              & $0.395_{-0.055}^{+0.069}$ & $0.34_{-0.08}^{+0.30}$ & ${0.00_{-0.00}^{+1.48}}$ \\
L06              & $0.286_{-0.064}^{+0.077}$ & $0.29_{-0.11}^{+0.18}$ & $0.77_{-0.77}^{+0.67}$   \\
L06($\Delta=500$)& $0.310_{-0.038}^{+0.041}$ & $0.33_{-0.10}^{+0.12}$ & $0.95_{-0.95}^{+0.53}$ \\
\noalign{\smallskip}\hline\hline
\end{tabular}
\end{center}
\tablecomments{0.9\textwidth}{The quoted constraints are obtained after marginalization of all nuisance parameters.}
\end{table}

\section{Results}

Using the method described above, we have constrained the cosmological information
preferred by the five \fgas samples. The best-fit parameter values at 1$\sigma$ CL
using corresponding $\eta_{\rm{obs}}(z)$ expressions are summarized in Table~2.
In Figs. 1--3, we plot the marginalized posterior probabilities of the reference
cosmology for each sample, taking $\Ok=0$ in the left panels, and $\Ok=1-\Om-\Ol$ in the right.

Note that the Union2 compilation of SNe Ia suggests a
$\Om=0.270\pm0.021~\rm{flat}~\Lambda$CDM universe. This is a relatively strong constraint
from direct observations. For the \fgas sample measured under certain reference cosmology,
the constrained results should reflect both this reference model as well as the cosmology
indicated by the SNe Ia observations. This is actually what our consistency test is designed for.

In Fig. 1, for A04 $\Lambda$CDM, its reference cosmology $(\Om=0.3,\Ol=0.7)$ is close to
the SNe Ia cosmology $(\Om=0.27,\Ol=0.73)$. They are all well consistent with the
constrained cosmology within 1$\sigma$ CL (Panel (a)). However, the 1-dimensional analysis
result of A04 SCDM is not so good. The best-fit parameter is $\Om=0.545$, which deviates
from both the reference cosmology $(\Om=1,\Ol=0)$ and the SNe Ia cosmology. Such a result
is reasonable, because the reference cosmology and the SNe Ia cosmology themselves are
quite different. Nevertheless from the 2-dimensional analysis, the correct
reference cosmological information ($\Ol=0$) is unambiguously revealed by the best-fit
parameter value (Fig.~1b), which is a convincible evidence that our method can shed
light upon the intrinsic reference cosmology of \fgas measurement. Generally speaking,
using the datasets reported by \citet{2004MNRAS.353..457A}, we proved the validity of our
method.

The analysis results of L06 and L06($\Delta=500$) are displayed in Figs. 2 and 3.
For both L06 (using priors/allowances proposed by Allen et al.) and L06($\Delta=500$)
(using priors/allowances proposed by Ettori et al.), the constrained cosmological
parameters are always consistent with its reported reference cosmology
$(\Om=0.3,\Ol=0.7)$ within 1 $\sigma$ CL. The priors on nuisance parameters proposed by
Ettori et al. in modeling \fgas, is rather strong compared with those proposed by Allen
et al.. Therefore one must be exceedingly careful when trying to derive cosmological
constraints via \fgas results using those stringent assumptions. Comparing Fig. 2 with Fig.
3, it is also clear that the CLs are enlarged owing to the change of $\eta_{\rm{obs}}(z)$
from Eq. (\ref{eta_obs04}) to Eq. (\ref{eta_obs08}). This is reasonable since
Eq. (\ref{eta_obs08}) includes more nuisance parameters, which are capable of reflecting
more physical effects and systematic uncertainties, and thus is a more generalized
expression.

However, as shown in Figs. 2 and 3, the cosmological parameters preferred by E09 deviate
greatly from its reported reference cosmology $(\Om=0.3,\Ol=0.7)$, which can never be accommodated within 1 $\sigma$ CL, regardless
of 1- or 2-dimensional constraints. The inconsistency can reach as notably as 6 $\sigma$ CL, with $\Om=0.439_{-0.023}^{+0.026}$ for
the flat $\Lambda$CDM cosmology under $\eta_{\rm{obs}}(z)$ Expression of Eq. (\ref{eta_obs04}).
If more nuisance parameters are considered in modeling \fgas, \ie
$\eta_{\rm{obs}}(z)$ altered from Eq. (\ref{eta_obs04}) to Eq. (\ref{eta_obs08}),
the confidence regions are enlarged as expected, yet E09's inconsistency still
exists at least at 1.7 $\sigma$ CL ($\Om=0.395_{-0.055}^{+0.069}$).
We also note that no matter which $\eta_{\rm{obs}}(z)$ expression is adopted,
the best-fit values from 2-dimensional analyses always
read $\Ol=0$ (see Table~2). In light of the result from A04 SCDM, we argue that the E09
\fgas data in nature prefers a cosmology without a dark energy
component\footnote{It is necessary to point out that the original study by
\citet{2003A&A...398..879E}, which is followed and updated by
\citet{2009A&A...501...61E}, did employ a reference cosmology of the Einstein-de Sitter
($\Om=1,\Ol=0$) universe.}, which can lead to biased cosmological parameter constraints
when this dataset is combined with probes that support concordance cosmology.

\begin{figure}\label{figure1}
\plottwo{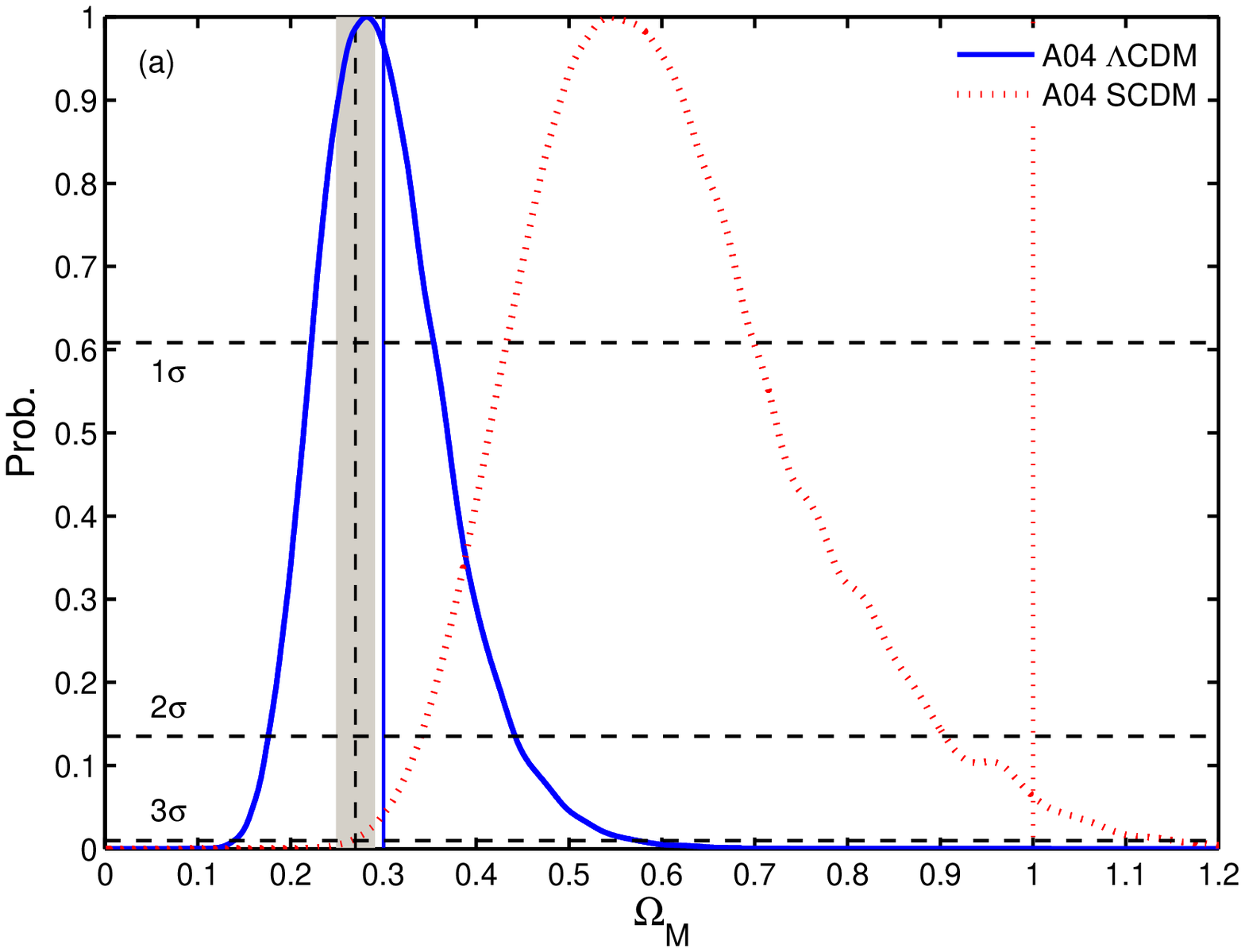} {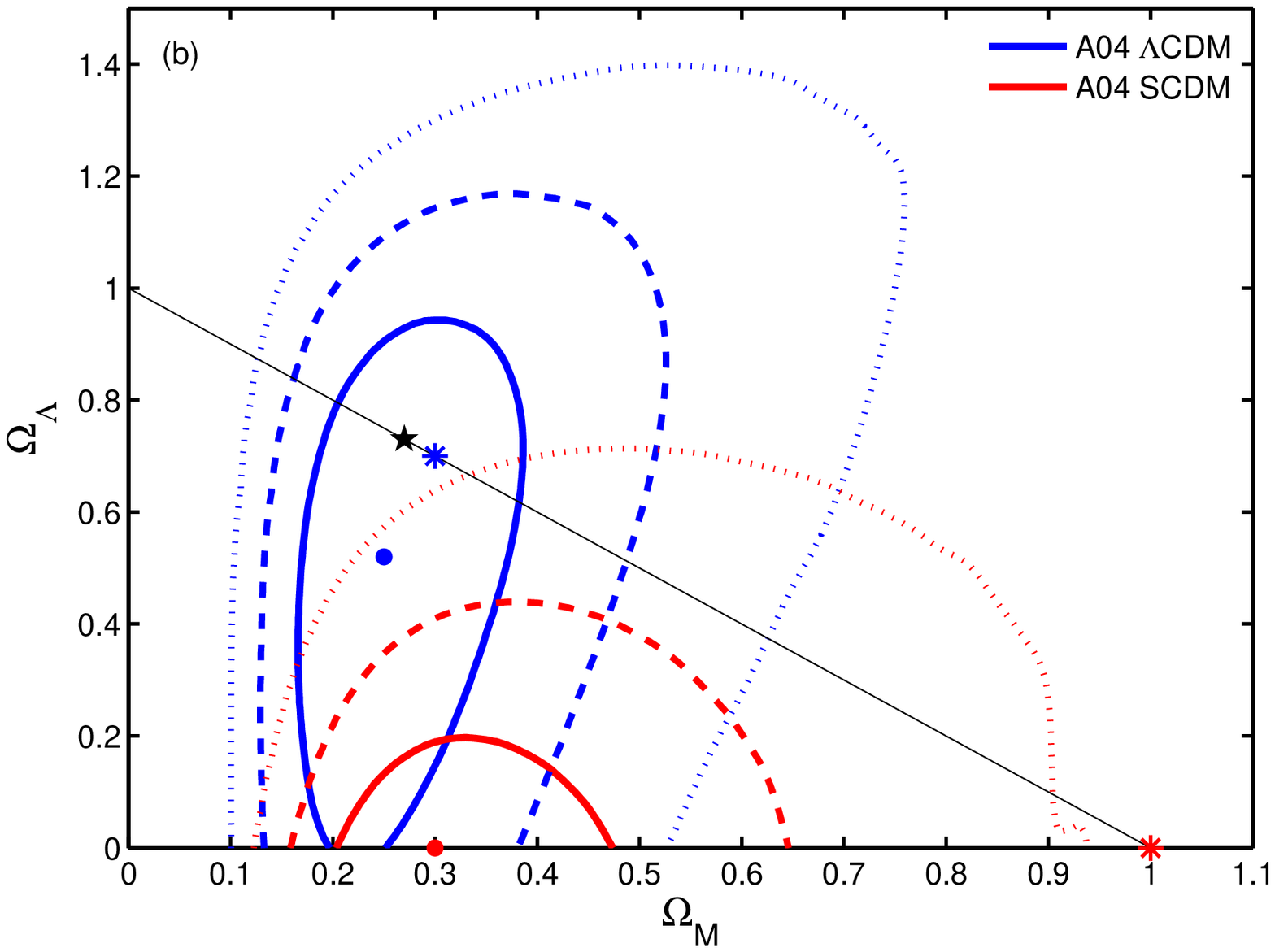}
\caption{Marginalized constraints on the preferred cosmological models by A04 $\Lambda$CDM and A04 SCDM, given the CDDR and Union2 SNe Ia data. Panel (a) shows the constraints on $\Om$, taking $\Ok=0$. The horizontal dashed lines correspond to 1, 2 and 3 $\sigma$ CLs respectively. The cosmological information from Union2 SNe Ia $(\Om=0.270\pm0.021)$ is marked by the vertical dashed line with the shaded region. The reported reference
cosmologies are indicated by the vertical solid and dotted lines, respectively. Panel (b)
shows the constraints in the $(\Om,\Ol)$ plane for a $\Lambda$CDM cosmology with $\Ok$
included as a free parameter. The 1, 2, and 3 $\sigma$ CLs are plotted by solid, dashed
and dotted lines, respectively. The best-fit values and reference cosmologies for these
samples are represented by big dots and stars in corresponding colors. The straight
thin line indicates a flat geometry. SNe Ia cosmology is marked by the pentagram.}
\end{figure}

\begin{figure}\label{figure2}
\plottwo{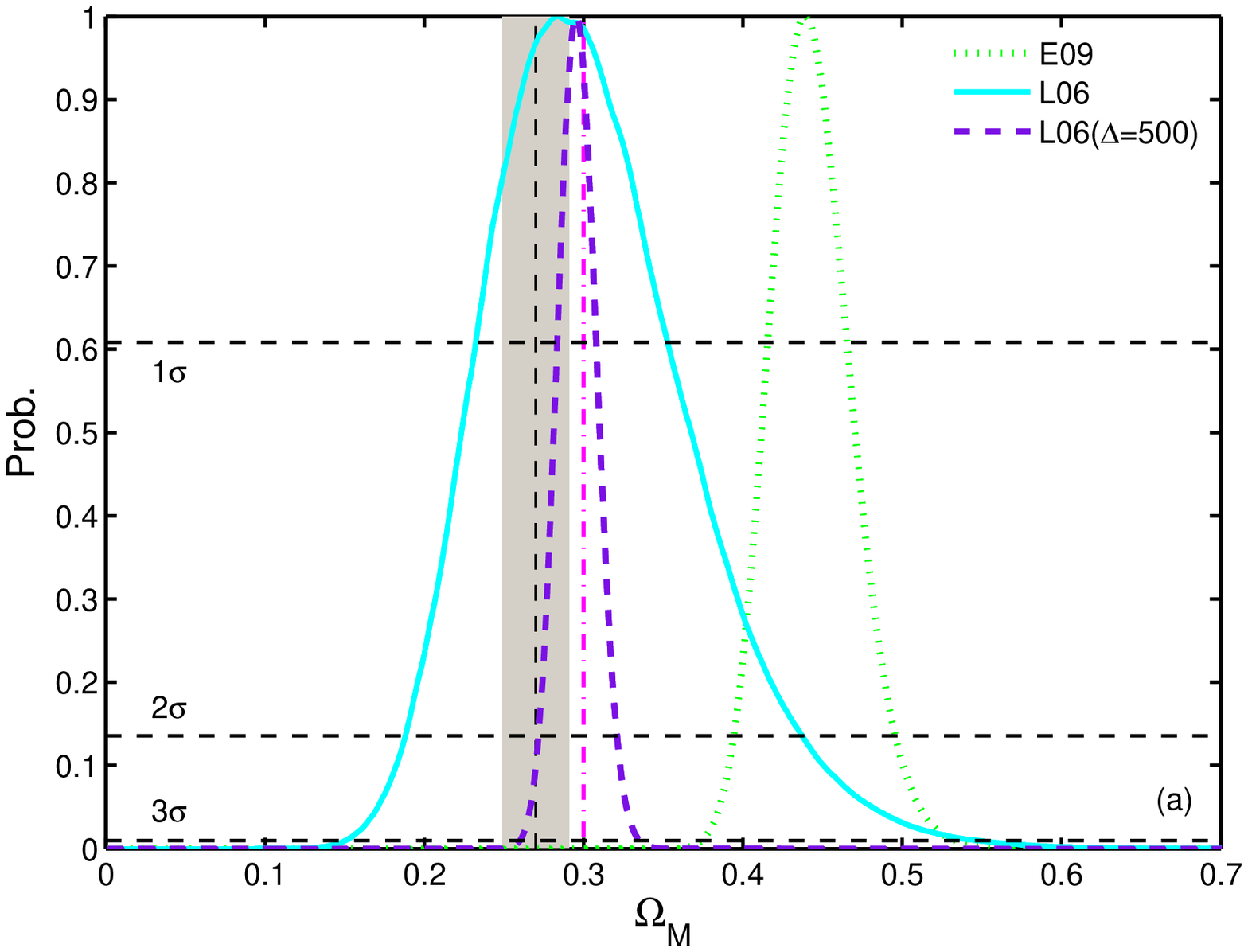}{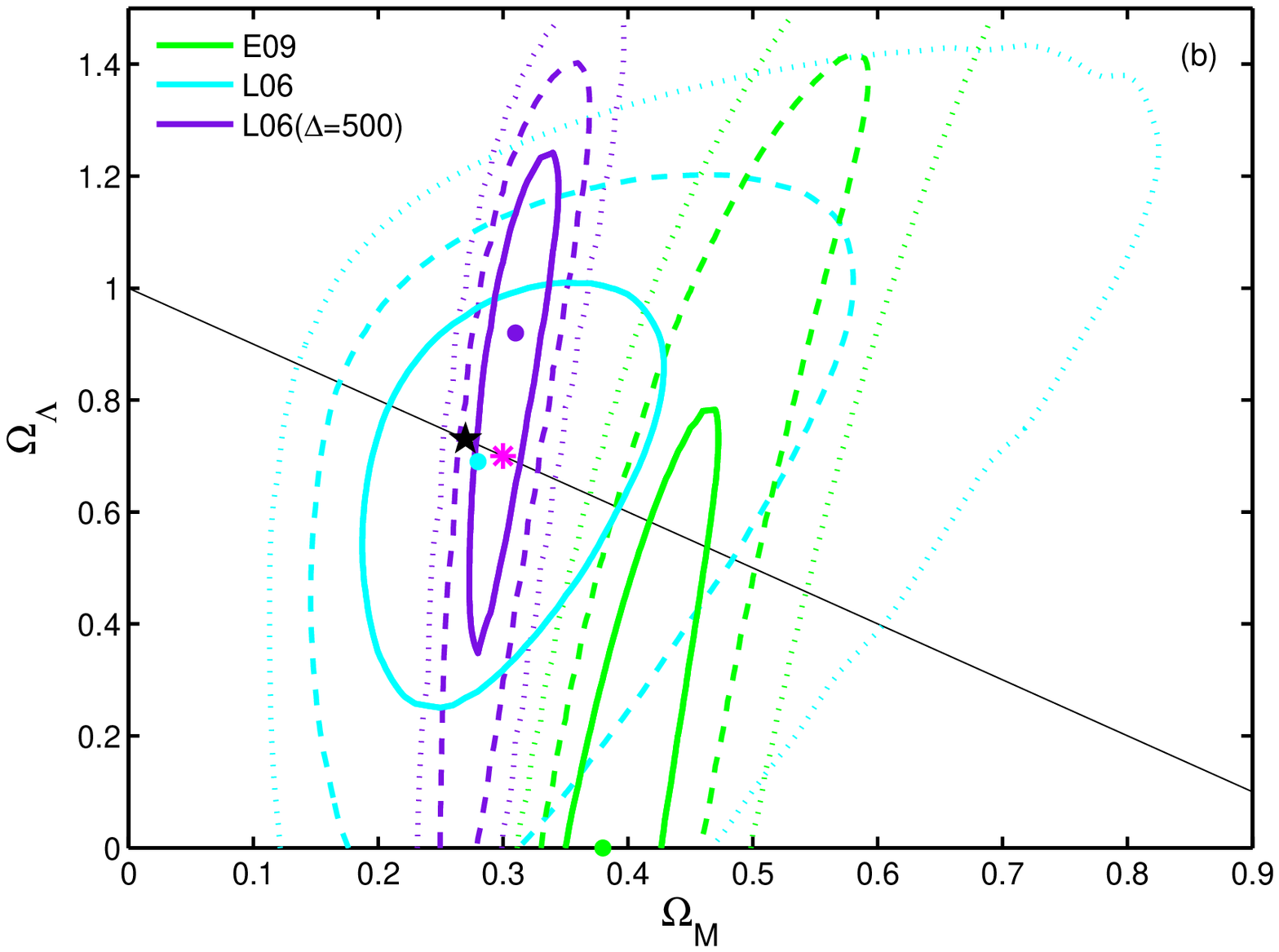}
\caption{Marginalized constraints on the preferred cosmological models by E09, L06 and
L06($\Delta=500$), given the CDDR and Union2 SNe Ia data. The $\eta_{\rm{obs}}(z)$ expression is given by Eq. (\ref{eta_obs04}). Panel (a) shows the constraints on $\Om$, under the assumption of a flat universe. The horizontal dashed lines correspond to 1, 2 and 3 $\sigma$ CLs respectively. The cosmological information from Union2 SNe Ia $(\Om=0.270\pm0.021)$ is marked by the vertical dashed line with the shaded region. These three samples' reference cosmological model is represented by the vertical
dash-dotted line. Panel (b) shows the constraints in the $(\Om,\Ol)$ plane for a usual $\Lambda$CDM cosmology model, with curvature kept free. The 1, 2 and 3 $\sigma$ CLs
are plotted by straight, dashed and dotted lines, respectively. The straight thin line
indicates a flat geometry. SNe Ia cosmology is marked by the pentagram. The best-fit values and reference cosmology for these samples are represented by big dots in corresponding colors and the star in magenta. Note that these samples have the identical reported reference cosmology.}
\end{figure}

\begin{figure}\label{figure3}
\plottwo{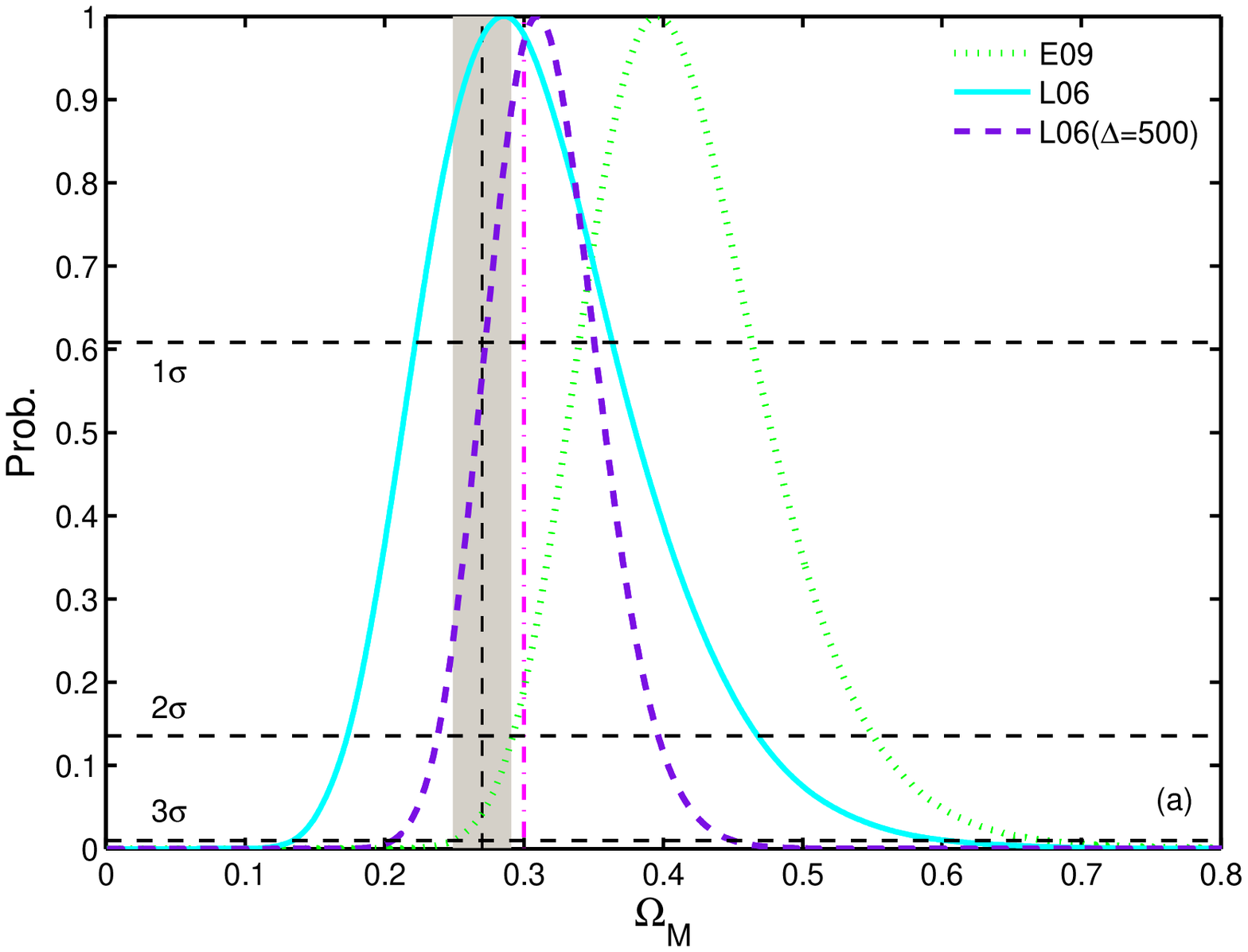}{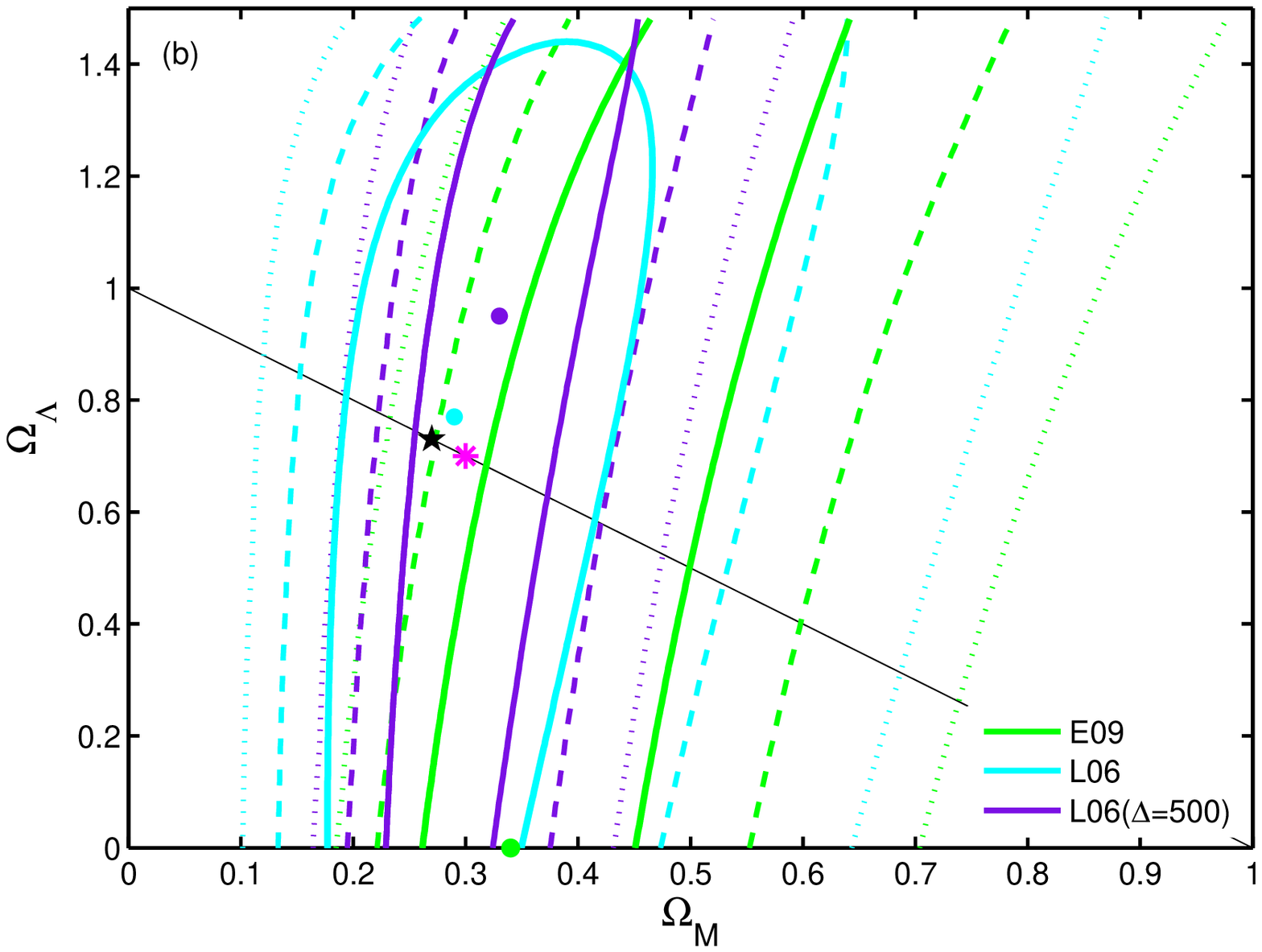}
\caption{Same as Fig. 2, except that the $\eta_{\rm{obs}}(z)$ expression is given by
Eq. (\ref{eta_obs08}).}
\end{figure}

\section{Conclusion and discussion}

In this paper, we proposed a consistency test to reveal the cosmological information preferred by X-ray \fgas measurements, using the CDDR and Union2 SNe Ia. We applied
this test to the \fgas samples provided by Allen \etal (A04 $\Lambda$CDM, A04 SCDM),
LaRoque \etal (L06, L06($\Delta=500$)) and Ettori \etal (E09). It is found that the samples of A04 $\Lambda$CDM, L06 and L06($\Delta=500$) show high level of consistency against our test. Despite the great discrepancy between the A04 SCDM's reference cosmology and the SNe Ia cosmology, our 2-dimensional analysis is still capable of probing its intrinsic cosmological information ($\Ol=0$) through the best-fit result.

However, our method reveals more than 3 $\sigma$ CL inconsistency for E09,
the \fgas dataset estimated by \citet{2009A&A...501...61E} assuming isothermal ICM.
Although endowed with an $\Om=0.3,\Ol=0.7$ $\Lambda$CDM reference cosmology as reported,
E09 shows special preference to an $\Ol=0$ cosmology. This result offers a reasonable
explanation for a recent CDDR test by \citet{2012MNRAS.420L..43G},
who found a significant conflict when using the \citet{2009A&A...501...61E} sample,
and this high-significance violation was only
spotted in \fgas sample from \citet{2009A&A...501...61E}\footnote{Note that the specific
datasets adopted by Gon{\c c}alves et al. and us from \citet{2009A&A...501...61E} are
different. Unlike the sample used by Gon{\c c}alves et al., our choice for investigation
(E09) obeys the assumption of isothermality, which plays a critical role in the
determination of $M_{\rm{tot}}$. Besides, we consider all available clusters data
for the purpose of minimizing systematic uncertainties.}.

The strength of nuisance parameters' priors proposed by Allen \etal and Ettori \etal is
also vividly demonstrated. The major differences between these two sets of priors exit in
the allowances on the depletion factor ($b$ or $b(z)$) and the baryonic matter density
($\Ob$)\footnote{The difference between allowances on $\xi$ is negligible, since it
affects \fgas values by less than ten percent \citep{2003A&A...398..879E}. Originally, the factor $A$ ($A=\left(\frac{H(z)\daz}{H^{\ast}(z)\das}\right)^{\xi}$) is introduced by \citet{2008MNRAS.383..879A} to account for the change in angle subtended by \r2500 as the underlying reference cosmology varies.}. The comparison between the results of L06 and L06($\Delta=500$) shows that the priors on these parameters given by Ettori \etal (at $\Delta=500$) are much more stringent than those given by Allen \etal (at $\Delta=2500$). However, since the X-ray background and the impact of ICM clumpiness can become a concern for $\Delta\leq500$ \citep{2011ARA&A..49..409A}, more reliable a priori knowledge on some influencing factors (baryon depletion, background contamination, cluster substructure, \etc) is still lacking for \fgas measurements and modeling at $\Delta=500$.

Furthermore, there are many physical processes affecting the measurements of
\fgas as well, particularly whether the cluster is in hydrostatic equilibrium or undergoes
major merger. Deviation from the equilibrium may give rise to large errors in \fgas results
\citep{2007ApJ...655...98N,2012ApJ...748..120D}, potentially leading to the inconsistency
presented by our analysis for the E09 sample.

In \citet{2009A&A...501...61E}'s study, the total baryon budget of clusters includes the contribution from the ICM gas and the cold baryons. The cold baryons themselves are composed of a stellar component and an intracluster light component. Additionally, their studies unexpectedly infer that there is still another baryonic matter component ($f_{\rm ob}$), whose percentage is non-negligible and can be as high as 25\%. This will bring significant systematic uncertainties to the measurement of the total baryon mass \citep{2006MNRAS.365.1021E}.

Moreover, another concern is the morphology hypothesis in modeling the ICM gas distribution.
Although our test proves the high consistency of the L06 sample, we still should bear in mind that the galaxy clusters in this sample are modeled under the spherical symmetry. Recently, the work by several groups \citep{lp,2012A&A...538A.131H} infer that comparing with the spherical geometry, the ellipsoidal morphology for the gas distribution is more preferable.

\begin{acknowledgements}
This work was supported by the National Basic Research Program of China (973 program, Grant
Nos. 2009CB824800 and 2012CB821804), the National Natural Science Foundation of China (Grant
Nos. 11033002 and 11173006), and the Fundamental Research Funds for the Central Universities.
\end{acknowledgements}


\end{document}